\documentclass[12pt]{article}
\pdfoutput=1
\usepackage{subfigure}
\usepackage{amssymb,amsmath}
\usepackage{graphicx}
\usepackage{color}
\usepackage[colorlinks=true
,urlcolor=blue
,citecolor=blue
,linkcolor=blue
,pagecolor=blue
,linktocpage=true
,pdfproducer=medialab
]{hyperref}
\usepackage[a4paper,width=15cm]{geometry}
\makeatletter \renewcommand{\@dotsep}{10000} \makeatother
\usepackage{appendix}

\newcommand{\beq}{\begin{equation}}
\newcommand{\eeq}{\end{equation}}
\newcommand{\bea}{\begin{eqnarray}}
\newcommand{\eea}{\end{eqnarray}}

\begin{document}

\vspace*{1cm}
\begin{center}

 {\Large\bf  Sum Over Histories: Discrete Step Interpretation 

 } \vspace{1cm}

{\large   Muhammad Adeel Ajaib\footnote{ E-mail: adeel@udel.edu}}

{\baselineskip 20pt \it
University of Delaware, Newark, DE 19716, USA  } \vspace{.5cm}

\vspace{1.5cm}
\end{center}


\begin{abstract}

We study the transition of a particle between two points such that the particle takes discrete spatial steps in this transition. We analyze how the sum over histories interpretation of quantum mechanics can be implemented in this scenario. We show that the Euclidean propagator of a free particle is recovered if the minimum space interval is of the order or greater than the De Broglie's wavelength of the particle. We also describe the statistical ensembles that model this transition. Furthermore, we discuss a possible extension of this model to 2-dimensions which serves as an example to extend it to any number of spatial dimensions.

\end{abstract}

\newpage

\section{Introduction}\label{intro}

Numerous experiments have exhibited the probabilistic nature of our Universe at the quantum scales.
A further insight into this probabilistic character was developed by Richard Feynman with the path integral formulation of quantum mechanics \cite{feynman-1948,feynman-1965}. In this formulation the transition of a particle between two points is described by taking into account all possible particle trajectories between these points. Each path contributes a phase $e^{i S/\hbar}$, where $S$ is the action for that path. The amplitude or the propagator is the sum of the contributions from all the trajectories. 

Inspired by the path integral formalism, we describe here a comparatively simplified approach that leads to the Euclidean propagator of a free particle. In addition, this model can provide further insights into motion of the particle at quantum scales. These insights include the limit on the step size for the particle's motion. We will show that the Euclidean propagator in this scenario is valid for step sizes of the order or greater than the de Broglie wavelength of the particle. We also describe the statistical ensembles that characterize this transition. A motivation to  introduce these ensembles is the relationship between the time evolution operator in quantum mechanics with the partition function as $Z=\mathrm{Tr}(e^{-\beta \hat{H}})$, through the transformation $\beta \rightarrow it/\hbar$.

The approach considered in this article overlaps with previous studies on the statistical approach to quantum mechanics. In reference \cite{Cruetz-1981}, for instance, Monte Carlo methods were used to evaluate the Euclidean path integral with individual paths of the particle defined on a discrete time lattice. These methods were then used to treat the anharmonic oscillator in 1 spatial dimension. Here, we shall focus on the motion of a free particle and derive the propagator using numerical techniques with Mathematica \cite{mathematica}. Moreover, we shall derive a limit on the minimum step size of the particle and extend our approach to more than one dimensions in the section \ref{model-2d}.

The paper is organized as follows: In section \ref{model-1d} we describe the model we employ in one dimension. We show that the sum over all paths leads to the Euclidean propagator and also present a possible definition of the probability for each path. We then describe the statistical mechanics model that mimics this transition. Section \ref{model-2d} generalizes the model to two and higher dimensions. We conclude in section \ref{conclude}.

\begin{figure}
\begin{center}
\includegraphics[scale=.5]{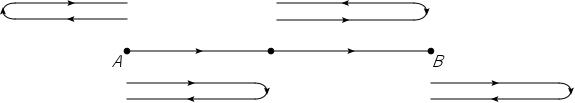}
\caption{The possible paths when the particle reaches point B in two steps by flipping once, i.e.  $W_{1d}(1,2)=4$. 
\label{fig1}}
\end{center}
\end{figure}

\section{Discrete steps in 1-D}\label{model-1d}

Consider the one dimensional motion of a free particle between two points A and B in discrete spatial steps of size $\delta x$. The particle covers a distance $d=x_A -x_B$ in time $\Delta t$. By discrete steps we assume that the particle `vanishes' at one point and appears at the other after time $\Delta t$. We will describe in subsection \ref{action-1d} what we mean by vanishing. Our aim is to choose $\delta x$ as small as possible in order to describe the continuous motion of the particle in good approximation. 
The particle starts at A and reaches point B after a minimum number of $m$ steps, i.e. $m=d/\delta x$. The particle can move from point A to B by taking all forward steps or can flip $j$ times as it moves towards B. Each subsequent path contains an additional flip. Therefore the number of steps in subsequent paths are given by
\begin{eqnarray}
N_0 &=& m \ \ \ \ \nonumber \\ 
N_1 &=& m+2 \nonumber \\ 
N_2 &=& m+6 \nonumber \\ 
. \nonumber \\ 
. \nonumber \\ 
N_j &=& m+2j,
\end{eqnarray}
where the $j$th trajectory has $m+2j$ steps. 
The number of possible trajectories the particle can take and the entropy for the $j$th path is given by
\begin{eqnarray}
W_{1d}(j,m) &=& \frac{N_j!}{n^{\uparrow}_j!  n^{\downarrow}_{j}!} 
\label{w-1d}\\
s_{1d}(j,m) &=& k_B \ln W_{1d} \ ,
\label{s-1d}
\end{eqnarray}
where $n^{\uparrow}_j=m+j$ represents the number of forward steps and $n^{\downarrow}_{j}=j$ represent the number of backward steps in the $j$th path. Note also that $n^{\uparrow}_j- n^{\downarrow}_{j}=m$ and $n^{\uparrow}_j + n^{\downarrow}_{j}=N_j$ so that the net displacement of particle is $m\delta x$ and $N_j \delta x$ is the total distance traveled. For example, with $j=1$ and $m=2$, $W_{1d}(1,2)=4$ implies that there are 4 possible trajectories the particle can take in which it flips once in each. These four paths are shown in Fig. \ref{fig1}. The arrows in the figure show the direction of motion and not the actual path of the particle. 

The number of possible paths $W_{1d}(j,m)$ increase as $j$ increases and so does the entropy $s_{1d}(j,m)$. However the ratio of entropy and the number of steps approaches a constant value $s_{1d}(j,m)/N_j \rightarrow k_B \ln 2$ as $j \rightarrow \infty$. The entropy is zero for the classical path.


\begin{figure}
\begin{center}
\includegraphics[scale=.45]{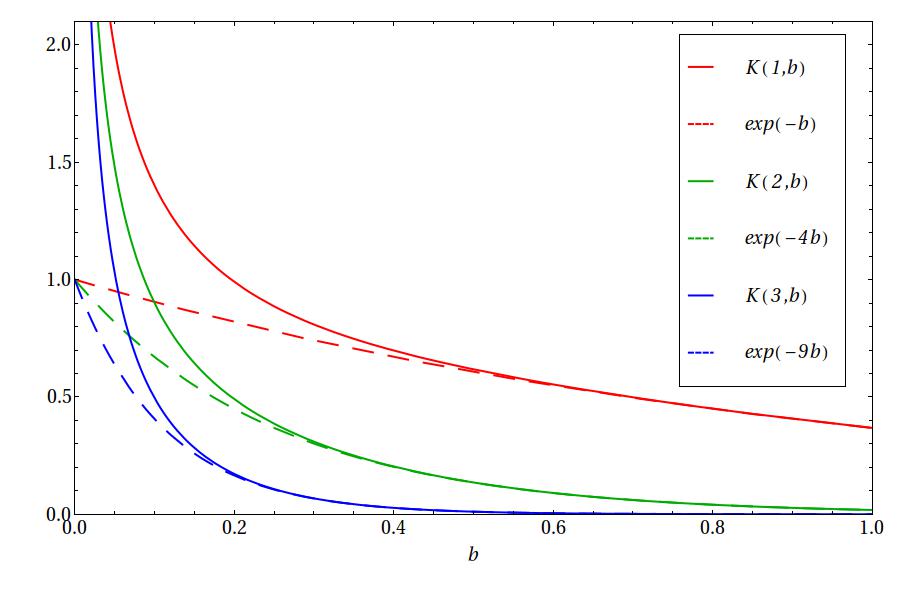}
\caption{Plot of the propagator given in equation (\ref{prop-1d}) as a function of the dimensionless parameter $b$ for different values of $m=1,2,3$. The thick colored lines show the numerical calculation of the sum over paths in (\ref{prop-1d}) using Mathematica whereas the dashed colored lines shows the decaying exponential function. We can see that the numerical sum approaches the exponential function exp$(-m^2 b)$ for $bm \gtrsim 0.5$ for each colored line. This implies that the numerical sum approaches the Euclidean propagator for $\delta x \gtrsim \hbar/Mv$.
\label{fig-propagator}}
\end{center}
\end{figure}

\subsection{Action and Propagator}\label{action-1d}
 
We assume that the particle moves with uniform velocity  $v_j$ in each of the possible paths. For the $j$th path the particle travels a distance $N_j \delta x$ in time $\Delta t$. We define the speed of the particle for the $j$th path as follows:
\begin{eqnarray}
v_j&=& \frac{N_j \delta x}{\Delta t} 
\end{eqnarray}
We further define the action for the $j$th path as 
\begin{eqnarray}
S_{1d}(j,m) &=& \frac{1}{2} M v_j^2 \Delta t  \nonumber \\
&=& \frac{M \delta x^2}{2\Delta t} N_j^2  \nonumber \\  
&=& c N_j^2  
\end{eqnarray}
where $c=M \delta x^2/2\Delta t$ and $M$ is the mass of the particle. The amplitude is given by summing the contributions of all the paths between the two points,
\begin{eqnarray}
K(m) &=& A \sum_{j=0}^{\infty} e^{-S_{1d}(j,m)/\hbar} = A \sum_{j=0}^{\infty} e^{-b(m+2j)^2}  
\label{prop-1d}
\end{eqnarray}
where $\hbar$ is the reduced Planck's constant and $b=c/\hbar$ is a dimensionless constant. We can calculate the sum in equation (\ref{prop-1d}) numerically using Mathematica \cite{mathematica} and Fig \ref{fig-propagator} shows the plot of this function for different values of $m$. We can see that for $b m \gtrsim 0.5 $ the sum approaches the limit $e^{-bm^2}$. The condition $b m \gtrsim 0.5 $ translates into the requirement $\delta x \gtrsim {\hbar}/{M v} \equiv \lambda$, where $\lambda$ is the de Broglie wavelength of the particle. We therefore obtain 

\begin{eqnarray}
K &=& A e^{-bm^2} \nonumber \\  
&=& A e^{\frac{-M \Delta x^2}{2\hbar\Delta t}}
\end{eqnarray}
only if 
\begin{eqnarray}
\delta x \gtrsim \frac{\hbar}{M v} 
\label{limit-1d}
\end{eqnarray}
where, $v=\Delta x/\Delta t$ and $\Delta x=m \delta x$. 
We choose $A=\sqrt{M/2 \pi \hbar t}$ to get the correct normalization of the propagator. Therefore, for a particle initially at $(x,t)=(0,0)$ the propagator is given by
\begin{eqnarray}
K(x,t) &=& \sqrt{\frac{M}{2 \pi \hbar t}} e^{\frac{-M x^2}{2\hbar  t}}
\end{eqnarray}
which is the solution of the heat equation. Wick rotating time $ t \rightarrow i  t$ yields the free particle propagator which is the solution of the Schrodinger's equation. Note that the sum in equation (\ref{prop-1d})  approaches infinity for $\delta x \rightarrow 0$ and the above propagator is not valid for $\delta x \lesssim \lambda$. The allowed range for $\delta x$ is therefore $\lambda \lesssim  \delta x \lesssim d$. Earlier we assumed that the particle vanishes at one point and appears at the next step. This limit on $\delta x$ sheds light on what we meant by `vanishing'. It means that at quantum scales this model describes the behavior of particles for $\delta x \gtrsim \lambda$. In other words we assume that at the quantum scale a more fundamental formalism describes the particle's motion for lengths less than the de Broglie wavelength of a particle.

\begin{figure}
\begin{center}
\includegraphics[scale=.5]{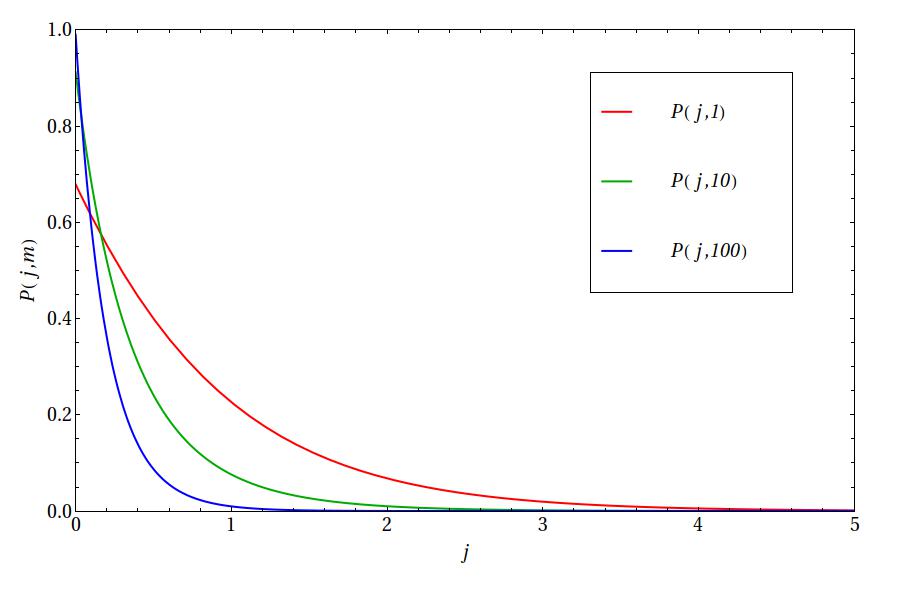}
\caption{Plot of the probability given in equation (\ref{prob-1d}) as a function of the number of times the particle can flip in a particular path. For a fixed step value $m=d/\delta x$ the probability decreases as $j$ increases. As the number of steps $m$ increases and the trajectory becomes more continuous the probability of the particle to follow the classical path increases. From equation (\ref{limit-1d}), the classical limit $\hbar \rightarrow 0$ implies the continuous limit  $m \rightarrow \infty$. For this limit the probability for the classical path is 1 and for all other paths is zero. For quantum scales with $\delta x_{min} \backsimeq \lambda$, the above plot shows that the probability of the particle to choose other paths is non-zero but small. Moreover, with $\delta x_{min}= \lambda$ the quantum effects are more prominent for smaller distances.
\label{fig-prob}}
\end{center}
\end{figure}

\subsection{Probability}\label{probability-1d}

 One possible way of parameterizing the probability of the particle to choose the $j$th path is as follows:
\begin{eqnarray}
P_{1d}(j,m)=  \frac{1/W_{1d}(j,m)}{\sum_{j=0}^{\infty}1/W_{1d}(j,m)}
\label{prob-1d} 
\end{eqnarray}
where the denominator normalizes the sum of the probability to 1. Note also that $1/W_{1d}(j,m)=e^{-s_{1d}/k_B}$. 
A drawback of defining the probability as in equation (\ref{prob-1d}) is that the probabilities for paths other than the classical paths are too small.  Note also that equation (\ref{prob-1d}) is not a unique way to parametrize the probabilities. Another possible way is given in equation (\ref{prob-1d-b}) of the appendix B.
Fig. (\ref{fig-prob}) shows the plot of the probability as a function of $j$ for different values of $m$. We can see from Fig. (\ref{fig-prob}) that for large values of $m$ the probability for the classical path approaches 1. This is because for large values of $m$ the path next to the classical path has a large multiplicity and therefore a small probability. In fact, for $m \rightarrow \infty$, the probability $P_{1d}(0,m)=1$ and $P_{1d}(j,m)=0 \ \forall \ j \ne 0$. In other words, as $m \rightarrow \infty$ ($\delta x \rightarrow 0$) the path next to the classical path has so much entropy that the particle does not traverse it. This implies that the limit $m \rightarrow \infty$  can be chosen as the classical limit. 

We described earlier that we need $\delta x$ to be as small as possible in order to describe the continuous motion of the particle in good approximation. As shown in  equation (\ref{limit-1d}), the Euclidean propagator is obtained if the lower bound on $\delta x=\lambda$. This is of course true for quantum scales but in order to describe classical objects we can choose $\delta x \rightarrow 0$ which is the limit $\hbar \rightarrow 0$. If we choose $\delta x=\lambda$ the probability for paths other than the classical path is small but non-zero. The fluctuations in the particle's trajectory get stronger for small distances.

The average distance, mean squared distance and the standard deviation for the $j$th path traveled by the particle, i.e.,
\begin{eqnarray}
\langle x_j \rangle  &=&  m \ \delta x \nonumber \\
\langle x_j^2 \rangle  &=&  N_j^2 \delta x^2 \nonumber \\
\Delta x^2_j  &=&  \frac{4j(m+j)}{m^2} \langle x_j \rangle^2
\label{distance-1d}
\end{eqnarray}
where $\delta x$ is the step size. In the above equation $\langle x_j \rangle=x_B-x_A$ is the displacement of the particle and $\langle x_j^2 \rangle$ is the total distance traveled by the particle in the $j$th trajectory. $\Delta x^2_j$ measures the deviation from the mean path and the paths with large deviations are less probable since the probability decreases sharply with $j$ according to our choice of the probability in equation (\ref{prob-1d}), also shown in Fig \ref{fig-prob}.

\subsection{The Statistical ensemble}\label{ensemble-1d}
This transition can be modeled on ensembles of 2 level spin systems. Each of our spin system is a canonical ensemble which contains $N_j$ number of localized spin 1/2 particles which are placed in a magnetic field. There are two possible energy states of each particle, $+E$ and $-E$. We have chosen a two level system because a spin up ($+E$) represents a step forward on $x$-axis and a spin down ($-E$) represents a step backward. All our ensembles will have the same net magnetic moment since this represents the displacement of the particle. The partition function of the $j$th ensemble is given by
\begin{eqnarray}
Z_{1d}= e^{\beta_{j} E}+e^{-\beta_{j} E} \ ,
\label{pf-1d}
\end{eqnarray}
where $\beta_j=1/k_B T_j$ and $k_B$ is the Boltzmann constant. The requirement that every subsequent ensemble has one additional spin up or down leads to the following parametrization of the probability and the  number of particles 
\begin{eqnarray}
p_j &=& \frac{m+j}{N_j} \nonumber \\
q_j&=&\frac{j}{N_j} \nonumber \\
n^{\uparrow}_{j} &=& m+j \nonumber \\
n^{\downarrow}_{j} &=& j
\label{ens-1d}
\end{eqnarray}
Here, $p_j$ ($q_j$) is the probability of spin up (spin down). Similarly, $n^{\uparrow}_{j}$ ($n^{\downarrow}_{j}$) is the number of spin up (spin down) particles in the $j$th ensemble. For a two level system with $N_j$ number of spins the average energy, mean square energy and the fluctuation in energy are given by 
\begin{eqnarray}
\langle E_j \rangle  &=&  m E  \nonumber \\
\langle E_j^2 \rangle   &=&  N_j^2 E^2 \nonumber \\
  \Delta E^2_j  &=&  \frac{4j(m+j)}{m^2} \langle E_j \rangle^2
\label{energy-1d}
\end{eqnarray}
These are similar to the expressions for distance of the particle in equations (\ref{distance-1d}). In the standard quantum mechanical picture the classical limit is recovered as $\hbar \rightarrow 0$. In our set of ensembles this limit corresponds to $k_B \rightarrow 0$. In this case the partition function would be $Z \simeq e^{\beta_j E}$ and this system represents the path with only forward steps which is the classical path. 
In our ensemble therefore $\hbar \rightarrow 0$ is equivalent to $k_B \rightarrow 0$.

\section{Two and Higher Dimensions}\label{model-2d}

\begin{figure}
\begin{center}
\includegraphics[scale=.45]{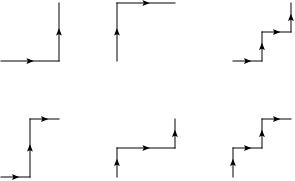}
\caption{Number of possible paths given by equation (\ref{w-2d}) for the particle to take for $j=k=0$ and $m_1=m_2=2$ ($W_{2d}=6$). All the paths shown are the minimum distance paths when the particle takes discrete steps. 
\label{fig-paths22}}
\end{center}
\end{figure}

\begin{figure}
\begin{center}
\includegraphics[scale=.5]{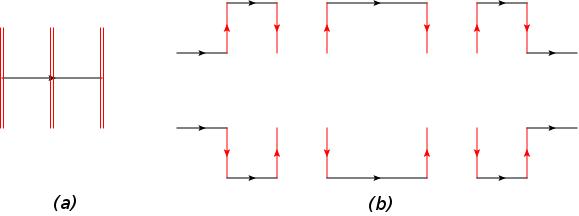}
\caption{Number of possible  particle trajectories for the path with one flip in the $y$-direction, i.e., $k=1, \ j=0$ and $m_1=2$ ($W_{2d}=12$). Fig 5(a) shows 6 paths together where the particle can flip (double red lines) in the $y$ direction at any of the points in the upward or downward direction. Fig 5(b) shows the other 6 paths the particle can take for this case.
\label{fig-pathk1}}
\end{center}
\end{figure}

In this section we extend our model in 1-D to two and higher dimensions. In two dimensions the method is somewhat different from the one we adopted in section \ref{model-1d}. We start with a 2D model which generalizes the ideas described in the previous section. We will first see that in order to represent the particle's motion we need to rotate the frame so that the particle's classical trajectory is on one of the axis. In this case the particle's motion will be along one direction with the fluctuations being in the perpendicular direction. Let us first assume that the particle moves from the origin to a point $(x,\ y)=(m_1 \delta x,\ m_2 \delta y)$ in the $xy$ plane. The multiplicity and entropy in this case are 
\begin{eqnarray}
W_{2d}(m_1,m_2) &=& \frac{N_{jk}!}{(n^{\uparrow}_{1j}! \  n^{\downarrow}_{1j} !)  (n^{\uparrow}_{2k} ! \ n^{\downarrow}_{2k} !)}  \nonumber \\
&=& \frac{(m_1 + m_2 +2j + 2k)!}{  (m_1+j)! \  j! \ (m_2+k)! \  k! }
\label{w-2d} 
\end{eqnarray}
\begin{eqnarray}
s_{2d} &=& k_B \ln W_{2d}
\label{s-2d}
\end{eqnarray}
where
\begin{eqnarray}
N_{jk}&=&n^{\uparrow}_{1j} +  n^{\downarrow}_{1j} + n^{\uparrow}_{2k} + n^{\downarrow}_{2k} \nonumber \\
&=& (m_1+2j)+(m_2+2k)
\end{eqnarray}
is the total number of steps the particle takes in the $x$ and $y$ directions. $W_{2d}$ quantifies the possible paths a particle can take. For instance, with $j=k=0$ and $m_1=m_2=2$, we get $W_{2d}=6$. These 6 paths are shown in Fig \ref{fig-paths22}. These paths are the minimum distance paths the particle can take. So, in 2 dimensions there are several routes that corresponds to the minimum distance of travel for the particle. We can see this from equation (\ref{w-2d}). For $j=k=0$ we get
\begin{eqnarray}
W_{2d}(m_1,m_2)=\frac{(m_1 + m_2)!}{m_1!  m_2!} 
\label{w-2d-0}
\end{eqnarray}
Therefore, there are several paths that corresponds to the minimum distance path. This is due to the fact that the particle can only take discrete steps. In more than one dimension therefore we will rotate our coordinate system so that the particle lies on the one of the  axis  (say $x'$-axis) of the new coordinate system. On rotating our coordinate system the new distance $x'$ of the particle from the origin will be our Euclidean distance in the original frame ($x'^2=x^2+y^2$) \cite{henning}. This implies that $m_2=0$ in our rotated coordinate system. In the rotated system, the distance traveled by the particle in path ($j,\ k$) with $N_j$ steps in the $x'$ direction and $2k$ steps in the $y'$ direction is $N_j \delta x + 2k \delta y$.
In this coordinate system the particle starts at $y'=0$. We therefore place $m_2=0$ in equation (\ref{w-2d}) and get
\begin{eqnarray}
W_{2d}(j,k,m_1)=\frac{(m_1+2j+2k)!}{(m_1+j)! j! (k!)^2} 
\label{w-2d-b}
\end{eqnarray}
$W_{2d}$ now quantifies the number of paths in the rotated frame. Fig. \ref{fig-pathk1} shows the possible paths for the particle when it flips once in the $y'$-direction ($k=1$). For the paths with $k=1$, $j=0$ and $m_1=2$ there are 12 equally probable trajectories for the particle.  We can parametrize the probability for this case as before:
\begin{eqnarray}
P_{2d}(j,k,m_1)= \frac{1/W_{2d}(j,k,m_1)}{\sum_{j=0}^{\infty} \sum_{k=0}^{\infty}1/W_{2d}(j,k,m_1)}
\label{prob-2d} 
\end{eqnarray}
The probability in this case also behaves similar to the 1 dimensional case. The difference is that sum now is over all the flips in the $x$ and $y$ directions. For example, for $(j,k,m_1)=(1,1,1)$, the probability of the path which include one flip in the $x$ and $y$ directions is $0.03 \%$. In this case the probability will be suppressed more since the number of possible paths have now increased. However, in the limit of $m_1 \rightarrow \infty$ the probability for all paths except the classical path is zero.

\subsection{Action and Propagator}\label{action-2d}
Next, following steps similar to section \ref{model-1d} we evaluate the propagator of the particle in two dimensions. The distance traveled by the particle in any step is $N_j \delta x + 2k \delta y$, where the $2k \delta y$ corresponds to the fluctuation in the $y'$ direction. The speed of the particle for the path ($j,\ k$) is:
\begin{eqnarray}
v_{j,k} &=& \frac{N_j \delta x+2k \delta y}{\Delta t}
\end{eqnarray}
As before, we define the action as
\begin{eqnarray}
S(j, k, m_1) &=& \frac{1}{2} M v_{j,k}^2 \Delta t  \nonumber \\
&=& \frac{M  \delta x^2}{2\Delta t} ( N_j+2k )^2  \nonumber   
\end{eqnarray}
where we have assumed $\delta x=\delta y$. The amplitude in this case is given by
\begin{eqnarray}
K(m_1) &=& \sum_{j=0}^{\infty}  \sum_{k=0}^{\infty} e^{-S(j,k,m_1)/\hbar} = \sum_{j=0}^{\infty}\sum_{k=0}^{\infty} e^{-b(m_1+2j+2k)^2}  
\label{prop-2d}
\end{eqnarray}
where $b=c/\hbar$. We can again calculate this sum numerically and obtain a similar limit on $\delta x$ as before. Therefore,
\begin{eqnarray}
K(x',t) &=& A e^{\frac{-M x'^2}{2\hbar\Delta t}} \nonumber \\  
&=& A e^{\frac{-M ( x^2+ y^2)}{2\hbar\Delta t}}
\end{eqnarray}
which is true if 
\begin{eqnarray}
\delta x &\gtrsim & \frac{\hbar}{M v} \nonumber   
\end{eqnarray}
where we have placed $ x'^2= x^2+ y^2$ to obtain the propagator in our original frame. We can use this strategy in any number of spatial dimensions. The constant $A$ can be chosen as in section \ref{action-1d}.

\subsection{The Statistical ensemble}\label{ensemble-2d}

In this case the ensemble is a system with two different types of particles, namely type $1$ and $2$. We assume that a mechanism maintains the two spins at different temperature, $\beta_{1j}$ and $\beta_{2k}$. In addition, a constraint is imposed on the system so that the number of spin up or down particles of a particular type are independent of the other. This is done so that this system represents the motion in two independent directions $x$ and $y$. At each location in the ensemble there can be either particle of type 1 or 2 with spin up or down. The partition function of the system is given by
\begin{eqnarray}
Z= e^{\beta_{1j} E_{1}}+e^{-\beta_{1j} E_{1}}+e^{\beta_{2k} E_{2}}+  e^{-\beta_{2k} E_{2}}
\label{pf-2d}
\end{eqnarray}
where $E_1$ and $E_2$ are energies of the two types of particles. We further impose the following restriction on the system described by the partition function in equation (\ref{pf-2d})
\begin{eqnarray}
\frac{ e^{\beta_{1j} E_{1}}+e^{-\beta_{1j} E_{1}}} { e^{\beta_{2k} E_{2}}+  e^{-\beta_{2k} E_{2}}}=
\frac{N_{1j}}{N_{2k}}
\label{rest-2d}
\end{eqnarray}
This condition renders the distribution of spins independent of each other. 
Again, this is to take into account the fact the motion of the particle along one axis is independent of the motion along the other. Using the above condition, the partition function in (\ref{pf-2d}) for $N_{jk}$ particles can be written as
\begin{eqnarray}
Z^{N_{jk}}= (e^{\beta_{1j} E_{1}}+e^{-\beta_{1j} E_{1}})^{N_{1j}} (e^{\beta_{2k} E_{2}}+  e^{-\beta_{2k} E_{2}})^{N_{2k}}
\left(1+\frac{N_{1j}}{N_{2k}}\right)^{N_{2k}}
\left(1+\frac{N_{2k}}{N_{1j}}\right)^{N_{1j}}
\label{pf-2d-b}
\end{eqnarray}
The average energy, mean square energy and the fluctuation in energy for each of the two spin types are given by equations similar to equations (\ref{energy-1d}). 
 The entropy of this system is given in equation (\ref{entropy-2d}) of the appendix and agrees with that obtained from (\ref{s-2d}) in the large $N$ limit. We next require that there are no spins of type 2 in the minimum entropy ensemble, i.e., $m_2=0$. This amounts to rotating the coordinate system. With $m_2=0$ the probability of the spin up or down is equal to 1/2 for the spin of type 2. Equation (\ref{pf-2d-b}) becomes
\begin{eqnarray}
Z^{N_{jk}}= (e^{\beta_{1j} E_{1}}+e^{-\beta_{1j} E_{1}})^{N_{1j}}  (e^{\beta_{2k} E_{2}}+  e^{-\beta_{2k} E_{2}})^{2k}
\left(1+\frac{N_{1j}}{2k}\right)^{2k}
\left(1+\frac{2k}{N_{1j}}\right)^{N_{1j}}
\label{pf-2d-b}
\end{eqnarray}
The above partition function represents the set of ensembles that model the transition of the particle in 2 dimensions with the rotated frame. 

\subsection{Extending to 3 or higher dimensions}
An approach similar to that described in previous sections can be used to describe the particle's transition in any number of dimensions. The number of steps in 3D will be $N_{jkl}=m_1+m_2 +m_3+2j+2k+2l$. We can similarly rotate the coordinate system so that the particle lies on one of the axis (say $x'$ axis, $m_2=0,\ m_3=0$). The multiplicity in this case will be
\begin{eqnarray}
W_{3d}(j,k,l,m_1)=\frac{(m_1+2j+2k+2l)!}{(m_1+j)!\ j! \ (k!)^2 \ (l!)^2} 
\label{w-2d-b}
\end{eqnarray}
where $l$ now is the number of steps in the $z$ direction. The probability can be defined as in equation (\ref{prob-2d}). The statistical ensemble would have a third type of spin and a condition similar to (\ref{rest-2d}) would render the number of spins independent of each other. Therefore, the analysis presented in these section can be used to extend the model to any number of spatial dimensions.

\section{Conclusion}\label{conclude}

We considered a simplified model of the transition of a particle between two points in discrete spatial steps. The aim was to study how the sum over histories or the path integral interpretation of quantum mechanics can be implemented in this scenario. This model is presented as an approximate description of the standard approach which involves integrating over all possible trajectories of the particle to derive the free particle propagator. 

We started with the particle's transition in one dimension and included trajectories in which the particle can flip any number of times. For each path we defined a multiplicity and thereby an entropy. 
We showed that the sum over all paths in this case leads to the Euclidean propagator for the particle if the step size is taken to be less than or of the order de Broglie's wavelength ($\lambda$). The propagator we obtained was valid for step size $\delta x \gtrsim \lambda$. In the classical limit ($\hbar \rightarrow 0$) the number of steps approach infinity and in this limit the classical path is the only contributing path. The probability of each path was taken to be proportional to the inverse of multiplicity. The paths involving large flips or fluctuations were shown to be less probable according to our definition. The classical path was defined to be the trajectory for which the number of steps $m \rightarrow \infty$ ($\delta x \rightarrow 0$). As $m \rightarrow \infty$ the path next to the classical path has so much entropy that the particle does not traverse it.

We then extended this description to two dimensions. Since there are several possible minimum distance paths for the particle in 2D we showed that this scenario can be described by rotating the frame such that the two points are on one of the axis of the rotated frame. The free particle  propagator can be derived for this case also leading to the same limit on the spatial step. We interpreted this limit on the spatial step as the limiting length to which this model is valid and below which a more fundamental formalism describes the particle's motion.

We also described the statistical ensembles that correspond to these transition. 
In 1D the ensemble is a two level system placed in a magnetic field. In 2D the ensemble is also a two level system but with two different types of spins. Additional constraint was imposed on this system so that the number of spins at any given temperature is independent of the other. This was to take into account the fact that the motion of the particle along one axis is independent of the motion along the other. For these ensembles the limit $k_B \rightarrow 0$ corresponds to the classical limit of the particle's transition. The ideas in 2 dimensions can be used to extend this model to any number of spatial dimensions. We have therefore demonstrated the possible implications of the Feynman path integral formulation of quantum mechanics in a model which describes the transition of a particle between two points in discrete spatial steps.

\section{Acknowledgments}
The author would like to thank Fariha Nasir and Muhammad Shahbaz for useful discussions.

\section*{Appendix A} 
\setcounter{equation}{0}  

\renewcommand{\theequation}{A-\arabic{equation}}

Here we present the expression for the entropy of the system of ensembles in 1 and 2 dimensions. For 1D, in the large $N$ limit, the entropy of a two level system described by the partition function in (\ref{pf-1d}) is given by
\begin{eqnarray}
s_{1d}=N_j k_B \left( \ln[\cosh(\beta_j E)]- \beta_j E \tanh(\beta_j E) \right)
\label{s-1d-a}
\end{eqnarray}
The condition that subsequent ensembles include one additional spin up and down translates to the following condition
\begin{eqnarray}
e^{ 2  \beta_j E} = \frac{m+j}{j}
\label{eq-ej} 
\end{eqnarray}
which implies
\begin{eqnarray}
\tanh \beta_j E &=& \frac{m}{N_j} \nonumber \\
\cosh \beta_j E &=& \frac{N_j}{2\sqrt{j(m+j)}}
\end{eqnarray}
Using above equations we can write the entropy of the two level in equation (\ref{s-1d}) system as
\begin{eqnarray}
s_{1d}= N_j \left[\ln \left( \frac{N_j}{\sqrt{j(m+j)}}\right)- \frac{1}{2} \frac{m}{N_j} \ln \left( \frac{m+j}{j}\right)\right]
\label{s-1d-b}
\end{eqnarray}
Similarly, for the system with two different types of spin 1/2 particles, the entropy is given by
\begin{eqnarray}
s_{2d}(j,k)&=& k_B \beta_{1j}  \langle E_1 \rangle+ k_B \beta_{2k}  \langle E_2 \rangle+ k_B  N_{jk}\ln Z_{2d}  \nonumber \\
s_{2d}(j,k)&=& s_{1 j}+ s_{2 k} +k_B \ln \left(1+\frac{N_{1j}}{N_{2k}}\right)^{N_{2k}}
+k_B \ln \left(1+\frac{N_{2k}}{N_{1j}}\right)^{N_{1j}}
\label{entropy-2d}
\end{eqnarray}
where $s_{1 j}$ and $s_{2 k}$ are the entropy of spins of type 1 and 2 given by expressions similar to (\ref{s-1d-b}). Note that in the above expression 1 and 2 represents the type of spin and we have suppressed the subscript 1d.

\section*{Appendix B} 
\setcounter{equation}{0}  

\renewcommand{\theequation}{B-\arabic{equation}}
The probability of the particle to choose the $j$th path can also be quantified as follows:
\begin{eqnarray}
P_{1d}(j,m)=  W_{1d}(j,m) \ 
p_j^{n^{\uparrow}_j} q_j^{n^{\downarrow}_{j}}
\label{prob-1d-b}  
\end{eqnarray}
where $W_{1d}$ is the number of possible ways a particular path can be traversed and is given in equation (\ref{w-1d}). Also, $p_j$ is the probability of a step forward and $q_j$ is the probability of a step backward. The other parameters can be parameterized as in equations (\ref{ens-1d})
\begin{eqnarray}
p_j &=& \frac{m+j}{N_j} \nonumber \\
q_j&=&\frac{j}{N_j} \nonumber \\
n^{\uparrow}_{j} &=& m+j \nonumber \\
n^{\downarrow}_{j} &=& j
\end{eqnarray}
The probability chosen in this manner decreases slowly with $j$ compared to equation (\ref{prob-1d}). A problem with this choice however is that the sum $\sum _0 ^{\infty} P_{1d}(j,m)$ of the probability is not convergent in 1D.


\begin{thebibliography}{99}

\bibitem{feynman-1948}
R. P. Feynman, Rev. Mod. Phys. 20, 367 (1948).


\bibitem{feynman-1965}
R. P. Feynman and A. R. Hibbs, \textit{Quantum Mechanics and Path Integrals} ~McGrawHill, New York (1965).

\bibitem{Cruetz-1981}
M. Creutz and B. Freedman, Ann. Phys. (N.Y.) 132, 427 (1981).

\bibitem{mathematica}
Wolfram Research, Inc., Mathematica, Version 8.0 (2011).

\bibitem{henning}
Henning F. Harmuth. \textit{Information Theory Applied to Space-time Physics, Chapter 3}, World Scientific (1992).

\end{thebibliography}
\end{document}